\documentclass[a4paper,twocolumn]{article}
\usepackage[T1]{fontenc}
\usepackage{amsmath,amssymb}
\usepackage{wasysym} 
\usepackage{graphicx}
\usepackage{authblk}

\title{Self-organized stationary patterns in networks of bistable chemical reactions}

\author{Nikos E. Kouvaris$^{*,a}$, Michael Sebek$^b$, Alexander S. Mikhailov$^c$, Istv\'{a}n Z. Kiss$^b$.}

\affil{
$^a$Department of Physics, University of Barcelona,  Mart\'i i Franqu\`es 1, 08028 Barcelona, Spain\\
$^b$Department of Chemistry, Saint Louis University, 3501 Laclede Ave., St. Louis, Missouri 63103, USA\\
$^c$Department of Physical Chemistry, Fritz Haber Institute of the Max Planck Society, Faradayweg 4-6, D-14195 Berlin, Germany\\
$^*$nikos.kouvaris@upf.edu}

\date{\today}
\begin{document}
\maketitle


\begin{abstract}
{\it
Experiments  with a network of bistable electrochemical reactions, organized in regular and non-regular tree networks, are presented to confirm an alternative to Turing mechanism for formation self-organized stationary patterns. The results show that the pattern formation can be described by identification of domains that can be activated individually or in combinations. The method was also demonstrated to localization of chemical reactions to network substructures and identification of critical  sites whose activation results in complete activation of the system. While the experiments were performed with a specific nickel electrodissolution system, they reproduce all the salient dynamical behavior of a general network model with a single nonlinearity parameter. This indicates that the considered pattern formation mechanism is very robust and similar behavior can thus be expected in other natural or engineered networked systems, which exhibit, at least locally, a tree-like structure.}
\end{abstract}

In biological context, many chemical reactions take place in discrete units, e.g., in cells, that form a complex network.\cite{VAR11,Newman:2003p6292} 
Similarly, recent advances in microfabrication allow the generation of engineered networks of reaction units, e.g., with lab-on-chip microdroplets,\cite{Toiya2008} electrode arrays,\cite{WIC13,Jia2012} BZ beads,\cite{TIN12} or microelectrofusion.\cite{Karlsson2002}
The interplays between local reaction kinetics (nodes), the physical processes that create coupling (link), and the architecture of the network in such systems can lead to a wealth of self-organized phenomena, including synchronization,\cite{WIC13,TIN12,Arenas2008} stationary Turing and oscillatory patterns,\cite{hor04,Nakao2010,Kouvaris2015a,Hata2014,Asllani2014b} or excitation waves.\cite{KOU14a,Isele2015,Steele2006} 

Stationary patterns generated via the Turing\cite{Turing1952} mechanism have been observed in experiments for both continuous\cite{Horvath:2009p6240} and networked\cite{Tompkins:2014:4397-4402} systems. 
Here, an alternative mechanism for emergence of stationary patterns in networks is experimentally explored. 
We focus on network-organized systems of bistable elements with diffusive connections between them. 
Bistable elements can be found in a broad class of chemical reactions (e.g., with autocatalysis\cite{chemistry_book,mik_book_I}) but also in cellular\cite{Graham2010} and engineered systems.\cite{Ikeda1980}
Such elements can have local or diffusive connections between them. 
For regular lattices and linear chains (i.e., for relatively simple networks), it is known that, under sufficiently weak coupling, the fronts fail to propagate, and thus stationary domains can be formed,\cite{Booth1992,Erneux1993,MIT98,Laplante1992} whereas at strong coupling the fronts spread\cite{Laplante1992,FLATGEN:1995p7306} and a uniform state is eventually established. 
Recently, analogues phenomena were theoretically investigated for complex networks and the formation of stationary domains, sensitive to the network topology, was predicted based on a simple model of regular trees and one-component bistable elements.\cite{KOU12,KOU13a} 
The aim of the present study is to experimentally identify the stationary pattern formation, induced by network structure, with chemical reactions in which autocatalysis can produce local bistable behavior.


Suppose that some substance can undergo chemical reactions in reactors occupying nodes $i$ ($i=1,\ldots,N$) of a network and this substance can spread diffusively from one node to another. Such a network-organized reaction-diffusion system is generally described by equations $\dot{u}_i = f(u_i) + K \sum_{j=1}^N\!\left(A_{ij}u_j-A_{ji}u_i\right)\!\,$, where $u_i$ is the chemical concentration in the node $i$, the function $f(u)$ specifies the local dynamics at the nodes, and the coefficient $K$ characterizes the strength of diffusive coupling. The network structure is determined by a symmetric adjacency matrix whose elements are $A_{ij}=1$, if there is a connection between nodes $i$ and $j$ ($i\neq j$), and $A_{ij}=0$ otherwise. Function $f(u)$ can be chosen in such a way that individual elements are bistable, e.g., with  autocatalysis (Supporting Information). For such models an approximate analytical theory is available for special networks representing regular trees with fixed branching ratio.\cite{KOU12}

\begin{figure}[th!]
\includegraphics[width=0.45\textwidth]{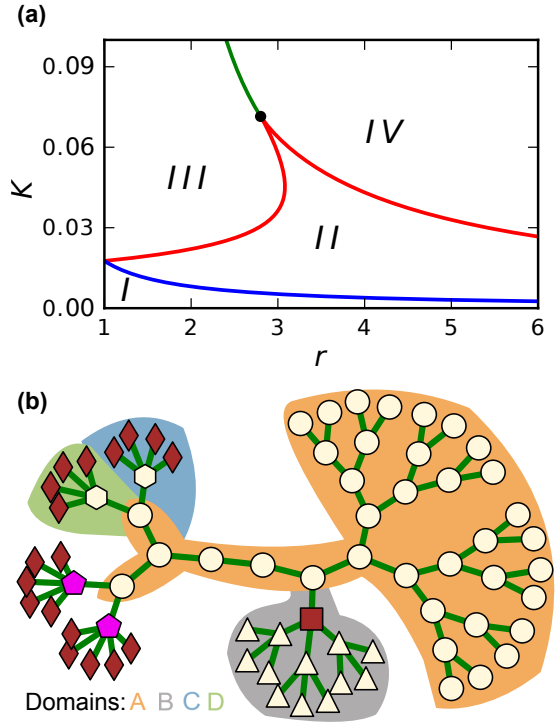}
\caption{a) The bifurcation diagram. Region $I$: Center and periphery activations are pinned. Region $II$: Center activation is pinned whereas periphery activation propagates towards the center of the tree. Region $III$: Center and periphery activations propagate in both directions, to the periphery and the center. Region $IV$: Center activation retreats whereas periphery activation propagates towards the center. b) Different activity domains in a non-regular tree network. Active domains $A$, $B$, $C$, and $D$, are distinguished by different colors. Activation of any node with the cream color ($\ocircle,\ \vartriangle,\ \varhexagon$) results in spreading of the activation over all nodes in the respective network domains $A$, $A\cup B$, $A\cup C$, or $A\cup D$. Activation of any node with the brown color ($\Square,\ \diamondsuit$) remains pinned on that node. Activation of a single node with the magenta color ($\pentagon$) retreats and vanishes resulting in the uniform passive state.}
\label{fig-bifurcation}
\end{figure} 

The analytical and numerical results are summarized in \figurename~\ref{fig-bifurcation}. Under given nonlinearity factors (Supporting Information), the two major parameters are the coupling strength $K$ and the branching ratio $r$. For weak coupling, the activation is pinned (i.e., it does not propagate) in region $I$. At the intermediate coupling strength there is a range of branching ratios (region $II$), where the center activation is pinned while the periphery activation propagates towards the center. At sufficiently strong coupling and relatively low branching ratios (region $III$), the center activation spreads towards to periphery, whereas the periphery activation propagates towards the center. At strong coupling and large branching ratios (region $IV$), the center activation retreats and the periphery activation propagates towards the center. The behavior of regular or non-regular tree networks can fully be interpreted in terms of these four regions. 

Figure~\ref{fig-bifurcation}b shows a non-regular tree network with branching ratios varying from one or two, three, four and five. Numerical simulations for this network were performed with the coupling constant $K = 0.04$, so that these domains corresponded to regions $II$, $III$ or $IV$ in the bifurcation diagram. Simulations were done by applying initial activation to one of the nodes and following the subsequent evolution until a stationary state was reached. Stationary patterns were then classified by identifying all the possible stationary states that could be obtained by such an initial activation. 

The developed patterns can be interpreted by the formation of domains activated individually or in combinations. The domains consist of groups of nodes that correspond to the same dynamical region and thus behave similarly. For example, domains of elements in region $III$ support spreading of activation towards both the center and the periphery. When such domains are adjacent to, or surrounded by nodes that are amenable to pinning or to retreating (e.g., in region $I$, $II$, or $IV$), the propagating fronts get pinned and stationary structures develop. Therefore, the observed structures consist of uniform domains separated by nodes amenable to pinning or to retreating. The exact configuration of the patterns depends on the architecture of the network, the applied coupling strength, and the nonlinearity of the reaction that altogether determine the assignment of the nodes to the different regions and the configuration of the domains. 

For the specific network, shown in \figurename~\ref{fig-bifurcation}b, activation of a node with the cream color resulted in spreading of activation to all nodes of the domains $A$, $A\cup B$, $A\cup C$, or $A\cup D$. Activation of a node with the brown color remained pinned on that node. Activation of a node with the magenta color retreated and vanished resulting in the uniform final passive state. Simulations were also performed by initially activating several network nodes, but essentially the same final patterns were then reached.


Experiments with networks of bistable electrochemical reactions were performed (Supporting Information). 
Each unit represented a corroding metal (nickel) wire that accommodated a complex reaction system (including, e.g., formation of multiple forms of metal oxides, bisulfate adsorption, oxygen evolution and metal dissolution) that exhibited bistable behavior. 
Moreover, coupling was established in the form of the charge flow between the wires (due to difference in electrode potential) which affected the rate of metal dissolution of the coupled electrodes.\cite{WIC13} 
Electrodes and external connections between them, correspond to the nodes and the links in all network diagrams. 

\begin{figure}[t!]
\includegraphics[width=0.45\textwidth]{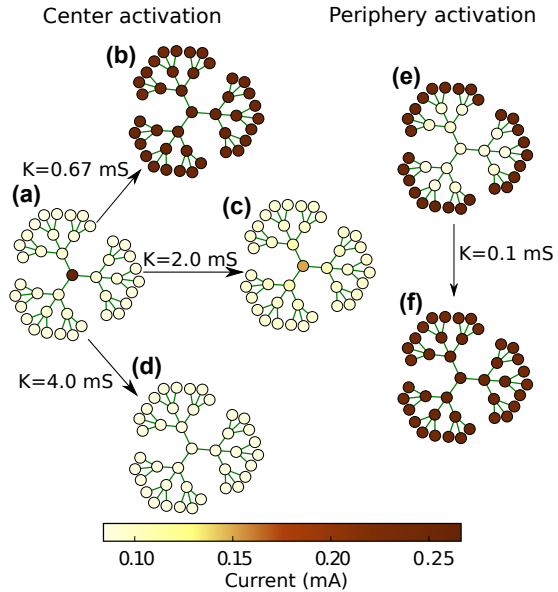}
\caption{Evolution of center and periphery activations on a regular tree with the branching ratio $r=3$ at different coupling strengths $K$. a) Initial condition of center activation. b) Spreading front, $K=0.67\ mS$; final state at $t=1008\ s$ is shown. c) Pinned front, $K=2.0\ mS$, $t = 2310\ s$. d) Retreating front, $K=4.0\ mS$, $t=68\ s$. e) Initial condition of periphery activation. If $K=0.04\ mS$ the front is pinned, $t=480\ s$. f) Spreading front, $K=0.10\ mS$, $t=464\ s$. $V = 1300 mV$; color coding indicated by the bar is used to display currents in network nodes.}
\label{fig-activation}
\end{figure}

First, experiments with regular trees were undertaken. Figure~\ref{fig-activation} shows that center activation in a four-layer tree with the branching ratio $3$ could result in spreading fronts (\figurename~\ref{fig-activation}b, region $III$) for weak coupling, pinned fronts (\figurename~\ref{fig-activation}c, region $II$) for moderate coupling, and retreating fronts (\figurename~\ref{fig-activation}d, region $IV$) for strong coupling. Similarly, periphery activation (\figurename~\ref{fig-activation}e) yielded either pinned fronts (\figurename~\ref{fig-activation}e, region $I$) at very weak coupling or spreading fronts (\figurename~\ref{fig-activation}f) at stronger coupling; such behavior was found in all regions II, III, or IV. 

\begin{figure}[t!]
\includegraphics[width=0.45\textwidth]{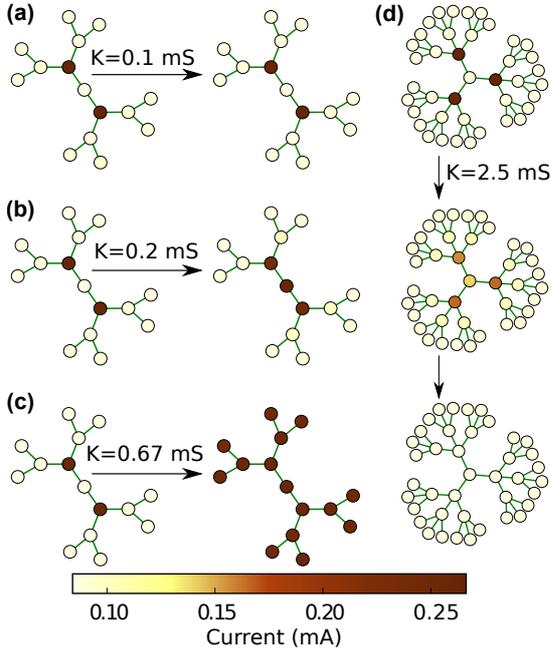}
\caption{a) Pinned fronts in both directions, $K=0.10\ mS$; final state at $t=402\ s$ is shown. b) Spreading front to the center and pinned front toward the periphery, $K=0.20\ mS$, at $t=602\ s$. c) Spreading fronts in both directions, $K=0.67\ mS$, at $t=538\ s$. d) A spreading front towards center and a retreating front from periphery, $K=2.5\ mS$. Middle snapshot at $t=25\ s$ and final state at $t=50\ s$. $V = 1300 mV$.}
\label{fig-regions}
\end{figure}

Figure~\ref{fig-regions} shows the evolution observed in the same network after activation of an intermediate node at different coupling strengths. In region $I$ (\figurename~\ref{fig-regions}a) the fronts are pinned on both sides. In region $II$ (\figurename~\ref{fig-regions}b) the center facing front spreads towards the center and the periphery facing front is pinned. In region $III$ fronts spread in both directions and finally activate the entire network (\figurename~\ref{fig-regions}c). In region $IV$ the center facing front propagates towards the center but the periphery facing front is retreated from the periphery finally establishing the passive state in the entire network (\figurename~\ref{fig-regions}d). All these experiments confirm the theoretical predictions for the regular tree networks.

\begin{figure}[t!]
\includegraphics[width=0.5\textwidth]{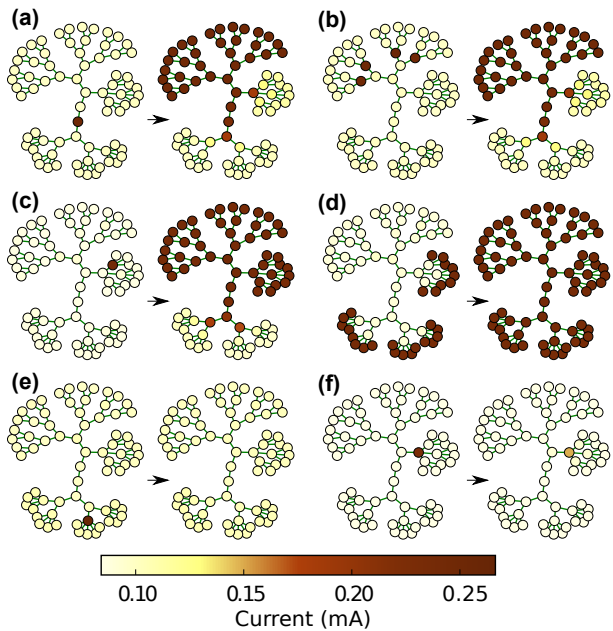}
\caption{a) Activation of a single node from the linear chain yields spreading fronts that activate the domain $A$ ($V= 1300\ mV$, final state at $t=654\ s$). b) Intermediate activation of four nodes in the domain $A$ with $r=2$ yields spreading fronts to activate the domain $A$ ($V=1295\ mV$, final state at $t= 241\ s$). c) Near periphery activation of a single node in the branch with $r=3$ yields spreading fronts which activate the domains $A \cup B$ ($V=1275\ mV$, final state at $t=336\ s$). d) Periphery activation of the branches with $r=3,\ 4$, and $5$ yields spreading fronts to complete network activation ($V=1280\ mV$, final state at $t=201\ s$). e) Activation of a single node in the branch with $r=5$ yields retreating fronts to complete network passivation ($V=1300\ mV$, final state at $t=54\ s$). f) Activation of a single more central node in the branch with $r=3$ yields pinned fronts ($V=1270\ mV$, final state at $t=801\ s$). All experiments performed at $K=1.3\ mS$.}
\label{fig-nonregular}
\end{figure}

Furthermore, we could also build the same non-regular tree as in the simulations in \figurename~\ref{fig-bifurcation}b. By performing experiments with regular trees under the same experimental setup, we could find (Supporting Figure, Figure S1) that at $K\approx1\ mS$ regions $II$, $III$, and $IV$ correspond to the branching ratios $r=3,\ 4$, $r=1,\ 2$, and $r\ge 5$, respectively. Similar to theoretical predictions, we could see that activations with a single (\figurename~\ref{fig-nonregular}a; Supporting Information, Video S1) or multiple nodes (\figurename~\ref{fig-nonregular}b; Supporting Information, Figure Video S2) within domain $A$ resulted in activation of all elements in domain $A$. When activation was applied to an intermediate element of domain $B$, pattern $A \cup B$ was observed (\figurename~\ref{fig-nonregular}c; Supporting Information, Video S3). To achieve complete activation of the network, peripheral nodes of the branches with the highest branching ratios $r=3,\ 4$ and $5$ had to be initially activated (\figurename~\ref{fig-nonregular}d; Supporting Information, Video S4). When a root node of a branch with high branching ratios $r=4$ or $r=5$ was initially activated, the activation could not spread and died out (\figurename~\ref{fig-nonregular}e). Furthermore, activation of more central node in the branch with $r=3$ resulted in a pinned front (\figurename~\ref{fig-nonregular}f).

Thus, we have experimentally demonstrated that bistable tree networks can support a rich variety of stationary patterns, determined both by the network architecture and initial activation conditions. Many features of the stationary patterns in a large,  non-regular network, can be predicted from the network topology based on experiments with four-layer trees that identify the dependence of the dynamical behavior as a function of the branching ratio. The experiments were performed with an electrochemical system; nonetheless, a surprisingly good agreement with theoretical predictions has been found despite the fact that the experimental system did not perfectly meet the idealizations made in the theory,\cite{KOU12} i.e., the state of a network element was described by more than a single variable and coupling between the elements was local, but not exactly of the diffusive form. This indicates that the found behavior is generic and robust; therefore it can be expected in various natural and engineered bistable networks. 

While only (regular or non-regular) tree networks were considered, our results are also relevant for large random networks that possess locally the tree structure.\cite{Barrat_book2008} 
Hence, formation of stationary domains should also be characteristic for random networks provided that such domains remain sufficiently small. Finally, it should be noted that, as has theoretically been shown,\cite{KOU13a} self-organized stationary domains can be controlled by introducing global feedbacks and experimental realization of such control would further facilitate the design of complex stationary structures on networks.


\paragraph*{\bf Acknowledgements.}
N.~E.~K. acknowledge financial support by the LASAGNE (Contract No.318132) EU project. M.~S. and I.~Z.~K. acknowledge support from National Science Foundation CHE-1465013.


\end{document}